\documentclass{amsart}

\usepackage{amsfonts}
\usepackage{amscd}
\usepackage{amsmath}
\usepackage{amssymb}
\usepackage{amsthm}
\usepackage{epsf}
\usepackage{latexsym}
\usepackage{verbatim}
\usepackage{color}
\input epsf.tex
\usepackage{bbm}
\usepackage{epstopdf}
\usepackage[pdftex]{graphicx}
\usepackage{fancyhdr}
\usepackage{wrapfig}

\theoremstyle{definition}

\newtheorem*{theorem*}{Theorem}

\usepackage{bold-extra}
\begin{document}

\title{A Game of Life  on Penrose Tilings}

\date{August 29, 2017}

\author[D. Bailey]{Duane A. Bailey}
\address{Department of Mathematics, Williams College, Williamstown, MA 01267, U.S.A.}
\email{bailey@cs.williams.edu}

\author[K. Lindsey]{Kathryn A. Lindsey}
\thanks{Second author supported by an N.S.F. Mathematical Sciences Research Postdoctoral Fellowship}
\address{Department of Mathematics, University of Chicago, Chicago,  IL 60637, U.S.A.}
\email{klindsey@math.uchicago.edu}


\begin{abstract}
We define rules for cellular automata played on quasiperiodic tilings of the plane arising from the multigrid method in such a way that these cellular automata are isomorphic to Conway's Game of Life.  Although these tilings are nonperiodic, determining the next state of each tile is a local computation, requiring only knowledge of the local structure of the tiling and the states of finitely many nearby tiles.  As an example, we show a version of a ``glider" moving through a region of a Penrose tiling.  This constitutes a potential theoretical framework for a method of executing computations in non-periodically structured substrates such as quasicrystals.  
\end{abstract}

\maketitle

\section{Introduction}

Inspired by Conway's Game of Life (\cite{Gardner}), various researchers have investigated properties of similarly-defined cellular automata played on Penrose tilings (e.g. \cite{Goucher, ImaiEtAl, CANDAR, OwensStepney2, OwensStepney}).  Since Penrose tilings are not periodic (\cite{GrunbaumShephard}), the challenge is to define a finite state cellular automaton which is locally computable -- meaning that the next state of a tile depends only on the current states of finitely many ``nearby" tiles -- and yet has interesting emergent global properties.  In particular, the existence of ``gliders" and the ability to support universal computation is of interest.

In this note, we present a natural method for embedding the original Game of Life in any tiling of the plane that arises via the multigrid method, a class of tilings that include Penrose rhomb tilings.  The resulting cellular automata, defined on quasiperiodic tilings, are locally computable and precisely mimic the behavior of the original Game of Life.  In particular, they admit gliders, signal delivery, universal computation, and reproduction.  

Quasiperiodic tilings may be thought of as ``toy models" of quasicrystals - physical substances which, like quasiperiodic tilings, exhibit order but not periodicity.  Penrose tilings, in particular, are a geometric model for icosahedral quasicrystals \cite{Mackay}.  This investigation is motivated by the promise of new media for hosting computation -- a topic which gives rise to the question of how to ``compute" in non-periodically structured substrates.  

\section{Statement of Results} \label{s:statementofresults}

\begin{figure}[ht]
\centering
\includegraphics{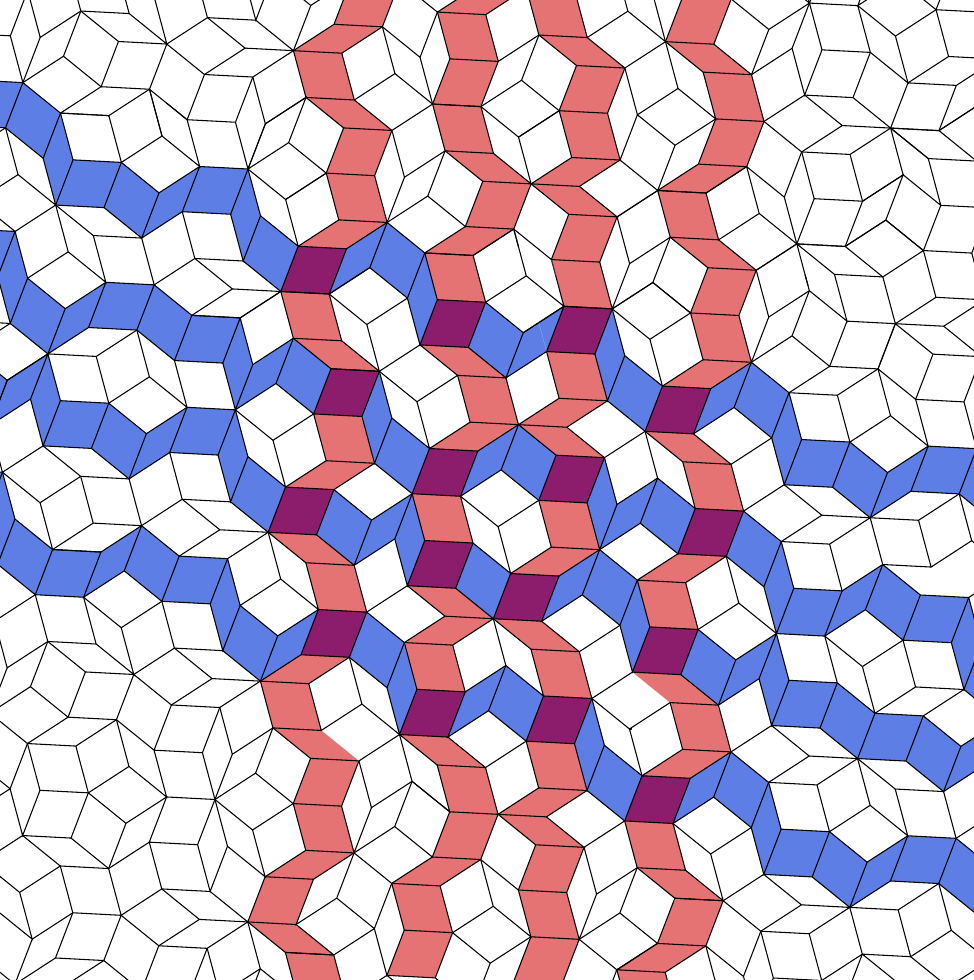}
\caption{Eight ribbons of a ``square grid graph" of ribbons in a Penrose tiling are colored in.  The two chosen vectors, $v_i$ and $v_j$, are the vectors parallel to the edges of the purple tiles.  The purple $(i,j)$-type tiles are the intersections of the grid of pink ribbons and the grid of blue ribbons.  Only four ribbons of each ribbon grid are colored, so that the reader may more clearly see the geometric structure, but one could complete the coloring by making all of the $(i,j)$-type tiles purple and extending pink and blue ribbons out from these tiles.}  
\label{f:ribbons}
\end{figure}

Quasiperiodic tilings generated by the ``multigrid" method admit \emph{ribbons} of tiles.   A ribbon consists of a bi-infinite sequence of sequentially adjacent tiles that share an edge that is parallel to some fixed vector.  In a tiling generated by an $n$-multigrid, every edge of the tiling is parallel to one of $n$ distinct vectors, say $v_1,\dots,v_n \in \mathbb{R}^2$.  Define a tile to be of \emph{type} $(i,j)$ if has two edges parallel to $v_i$ and two edges parallel to $v_j$.  A \emph{regular} tiling has ${n \choose 2}$ types of tiles, each of which is a parallelogram and has a unique type.  A tile with more than four edges, which by definition belongs to a singular tiling, may be of more than one type (a tile of type $(i,j)$ may have additional edges not parallel to $v_i$ or $v_j$.) \emph{For any choice of $(i,j)$, $i \not = j$, in any tiling arising from a multigrid construction, the set of all ribbons passing through all tiles of type $(i,j)$ forms a square grid graph} (i.e. the pattern formed by the ribbons is topologically the same as the pattern of lines on a piece of graph paper).  \emph{The ``vertices" of the square grid graph are the type $(i,j)$-tiles.}  See Figure \ref{f:ribbons}.  

This square grid graph structure determines a natural isomorphism between the set of type $(i,j)$ tiles in a regular quasiperiodic tiling arising from a multigrid construction and the grid of squares in Conway's original Game of Life.  This isomorphism preserves the property of adjacency of tiles within the lattice, and is unique up to choosing which tile represents the origin in the square grid $\mathbb{Z}^2$ and which directions of ribbons correspond to the positive $x$ and $y$ directions in $\mathbb{Z}^2$.   Since discerning ribbons only entails evaluating the parallelism of edges of adjacent tiles, picking out the natural lattice structure of the set of type $(i,j)$ tiles in any region requires only local knowledge of the tiling.

\emph{Our fundamental observation is that the original Game of Life may be ``played" on this lattice of type $(i,j)$ tiles.}  That is, in a regular tiling, each tile of type $(i,j)$ has precisely eight neighbors -- the eight $(i,j)$ type tiles that are ``neighboring" in the lattice structure determined by the square grid graph of ribbons.  Equipped with this observation, several obvious variants of the ``game" are possible:

\bigskip
\noindent {\bf Game Variants:}

\begin{figure}[ht]
  \begin{center}
    \includegraphics[width=0.23\linewidth]{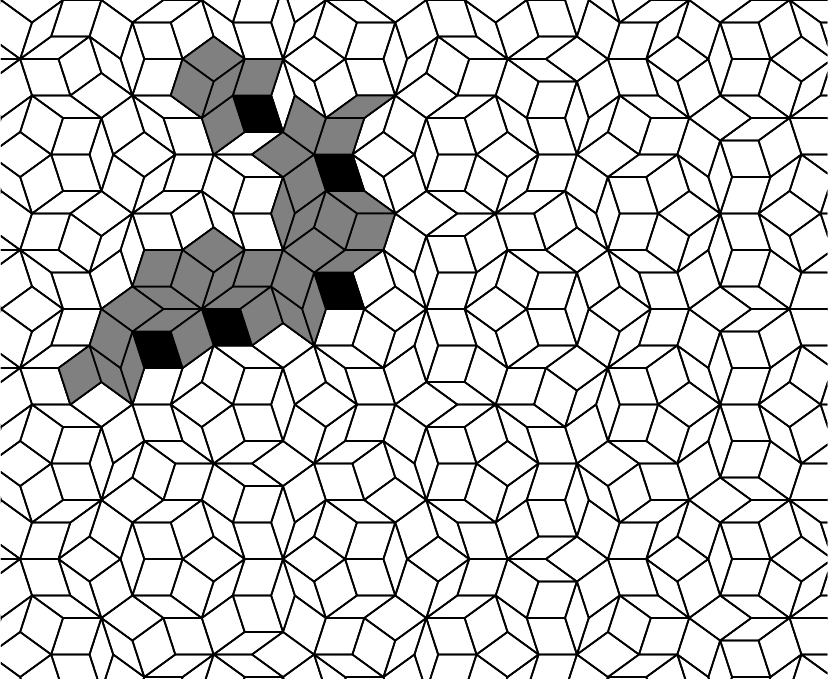} 
    \includegraphics[width=0.23\linewidth]{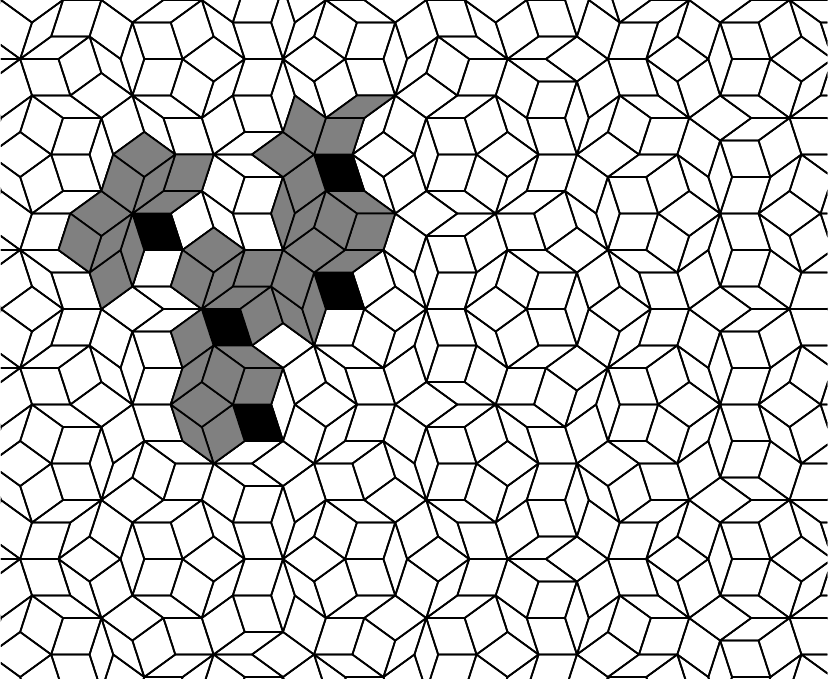}
    \includegraphics[width=0.23\linewidth]{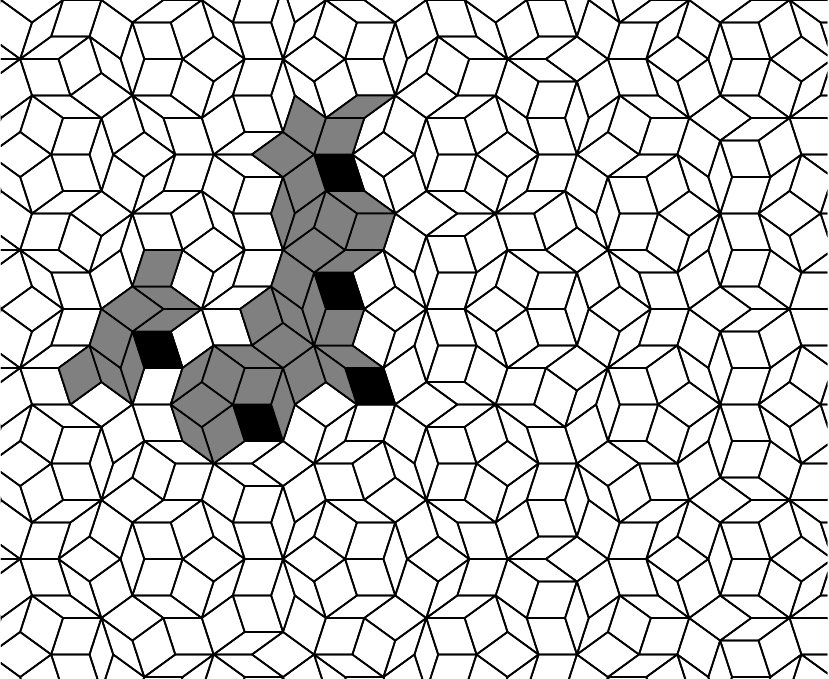}
    \includegraphics[width=0.23\linewidth]{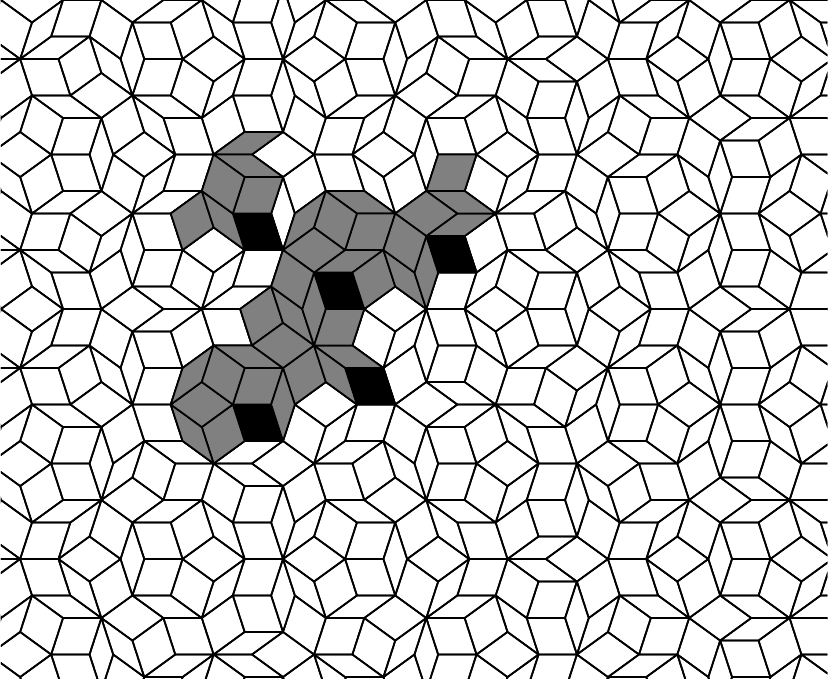} \vspace{.1cm} \\ 
    \includegraphics[width=0.23\linewidth]{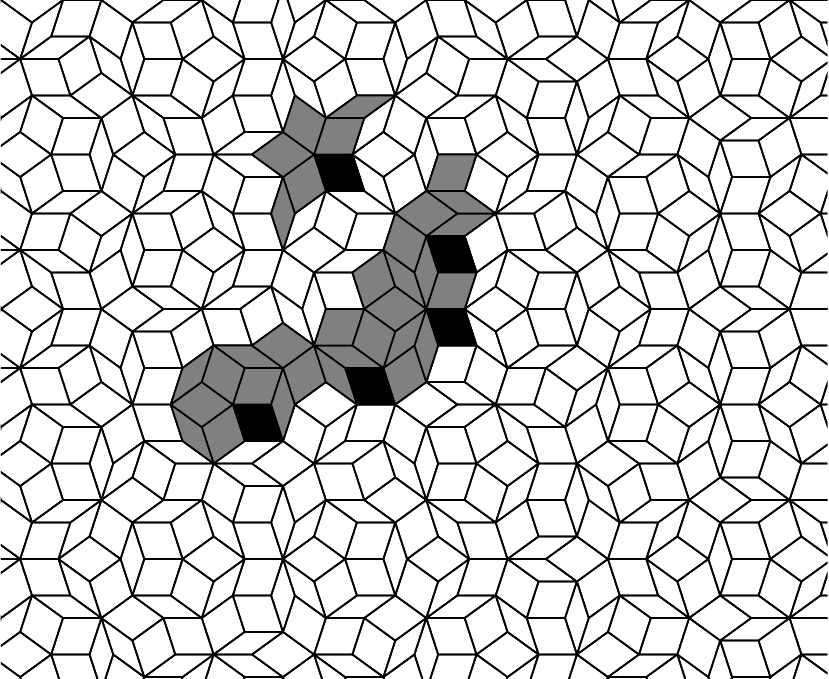}
    \includegraphics[width=0.23\linewidth]{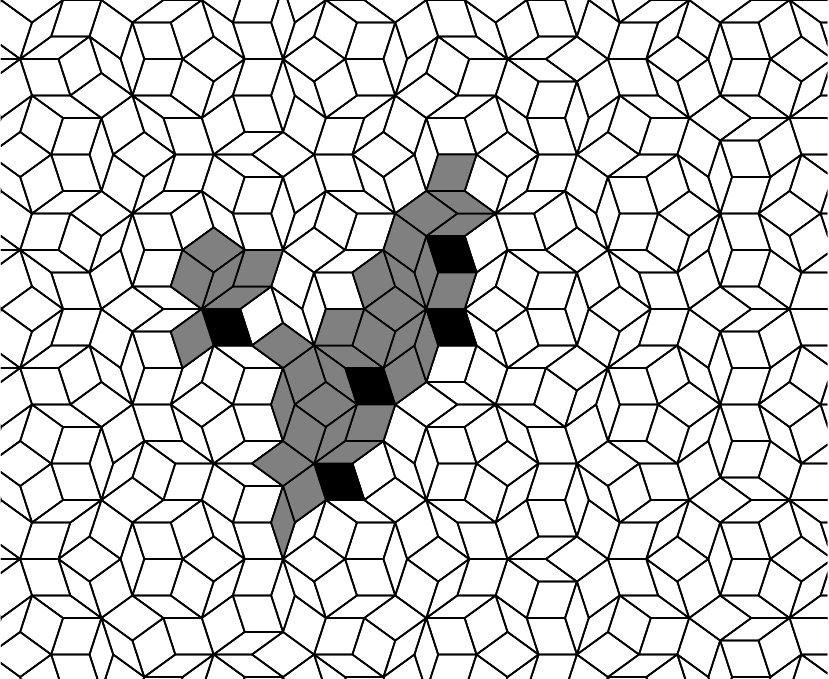}
    \includegraphics[width=0.23\linewidth]{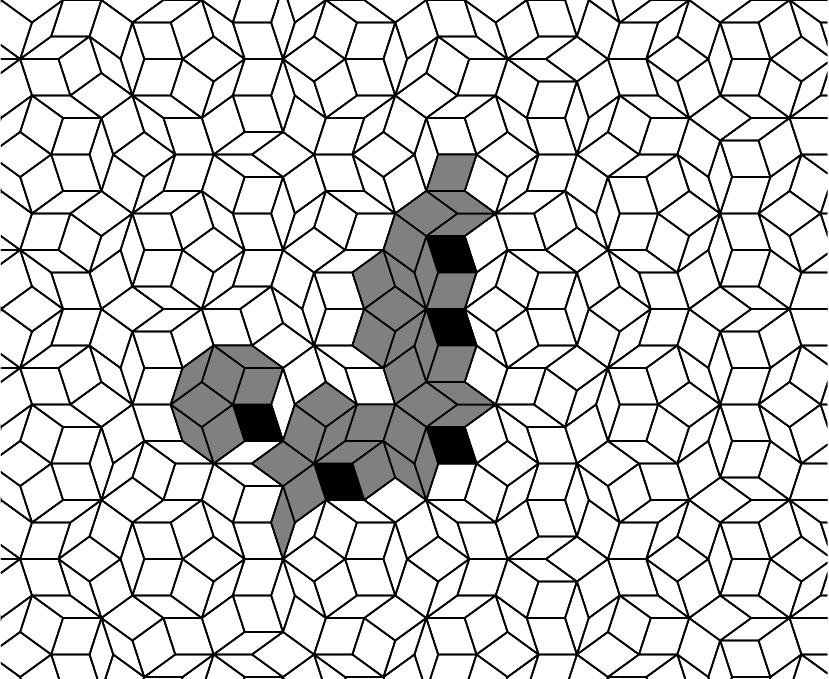}
    \includegraphics[width=0.23\linewidth]{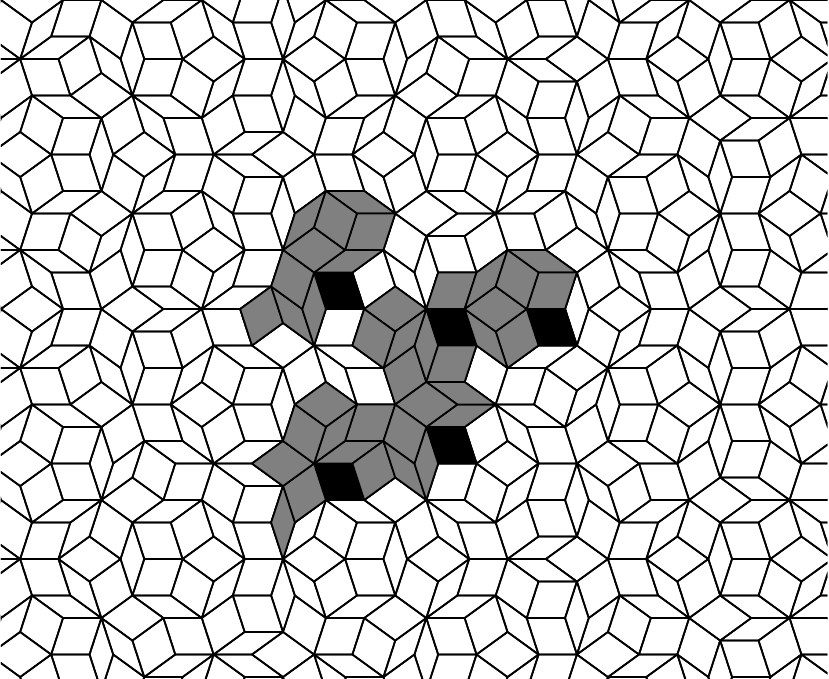} \vspace{.1cm} \\
    \includegraphics[width=0.23\linewidth]{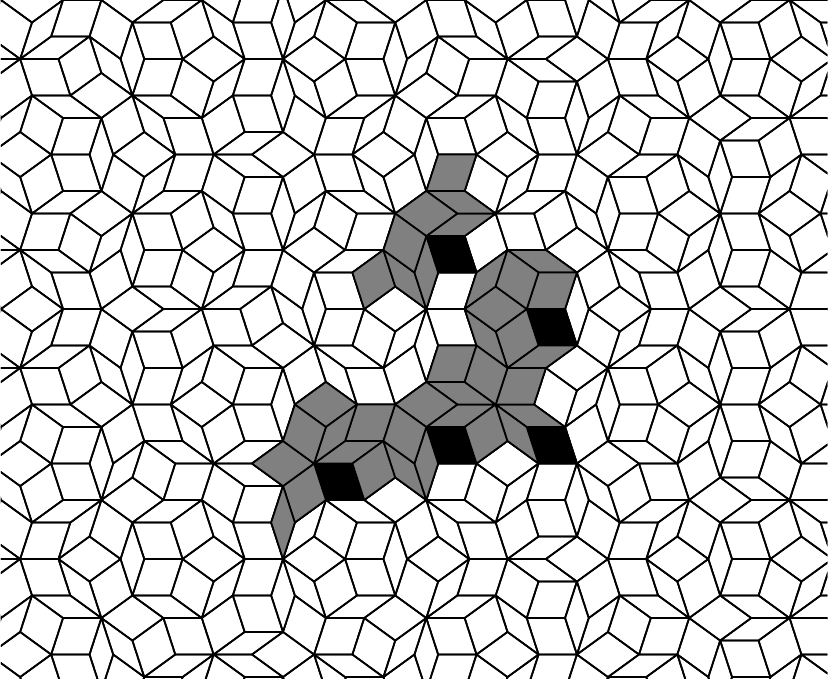}
    \includegraphics[width=0.23\linewidth]{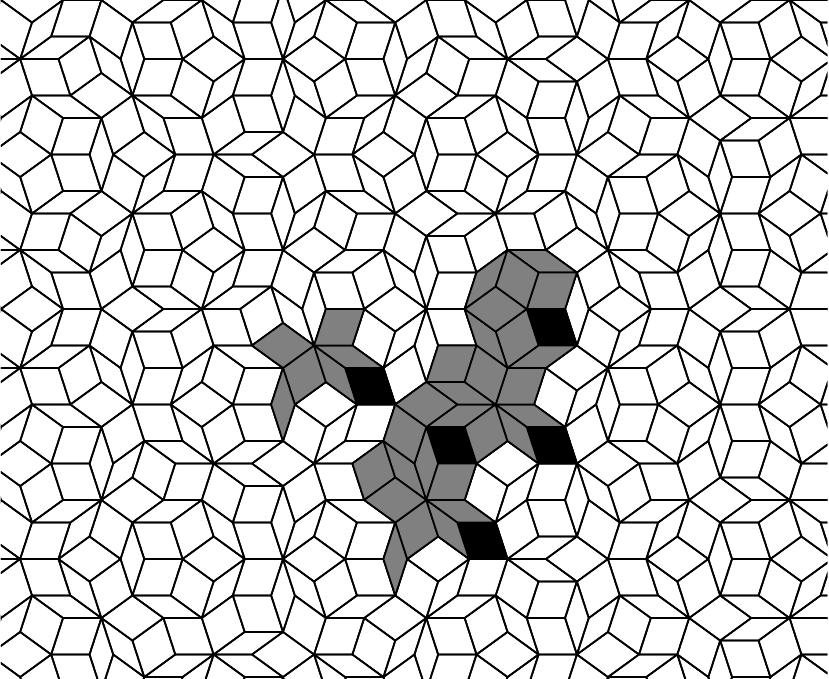}
    \includegraphics[width=0.23\linewidth]{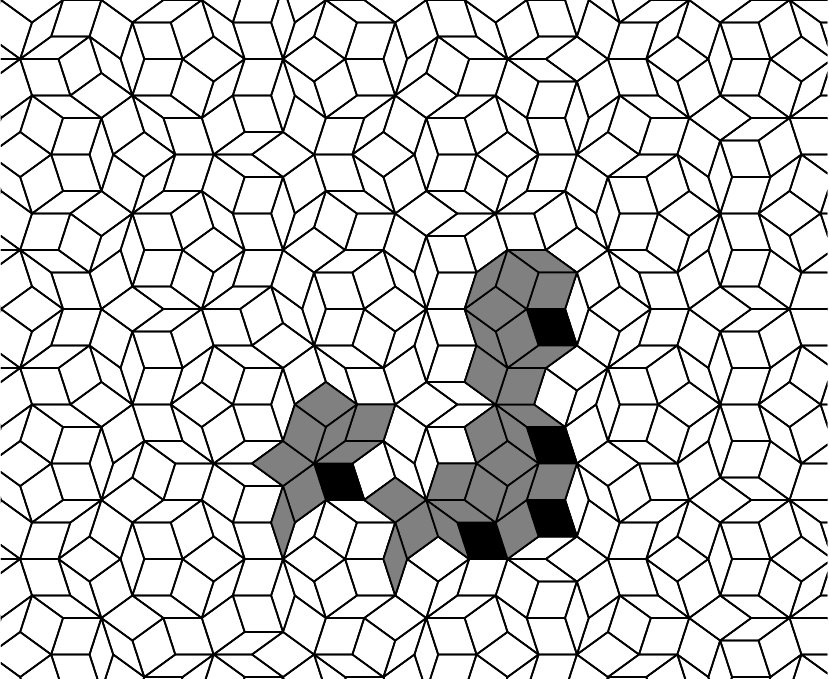}
    \includegraphics[width=0.23\linewidth]{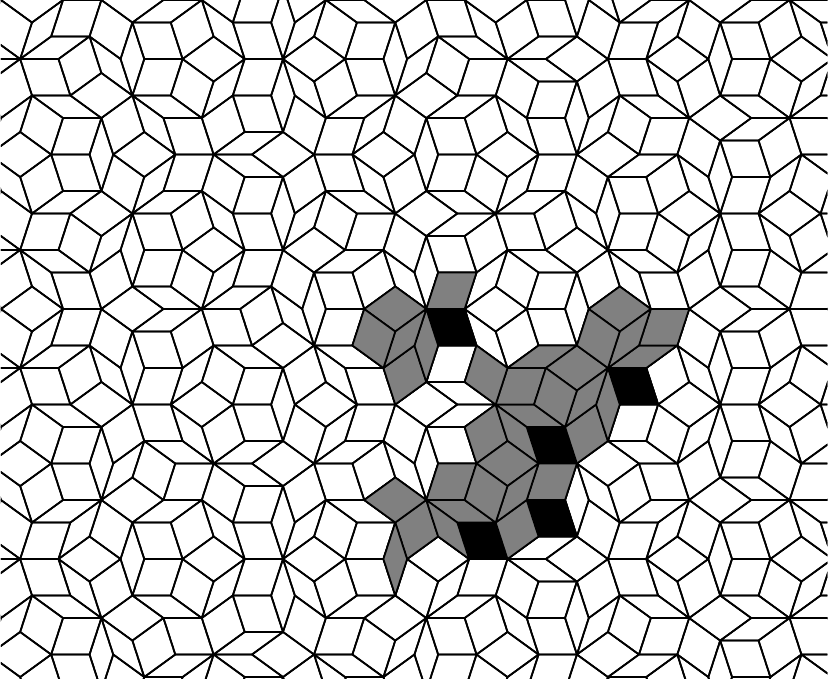} \vspace{.1cm} \\
    \includegraphics[width=0.23\linewidth]{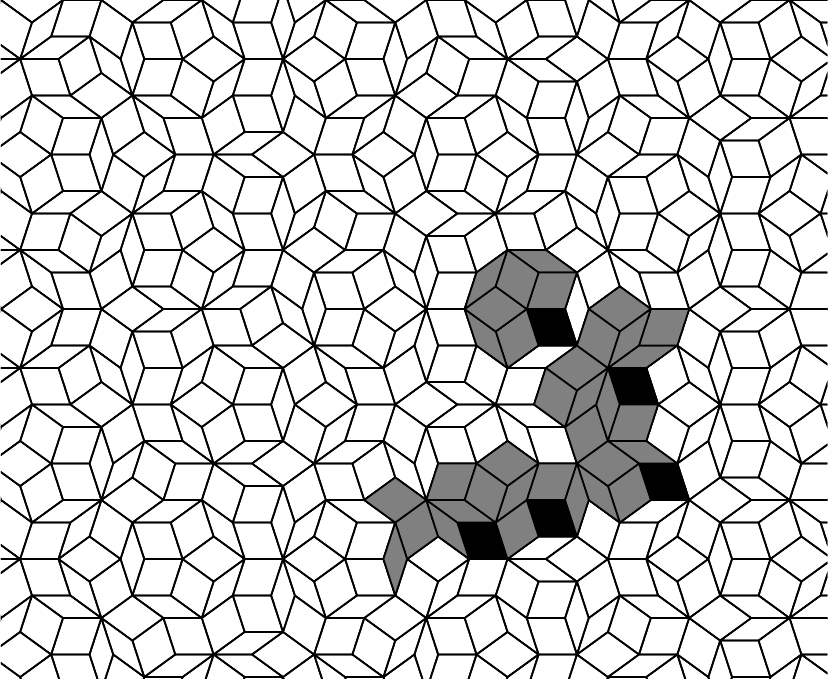}
    \includegraphics[width=0.23\linewidth]{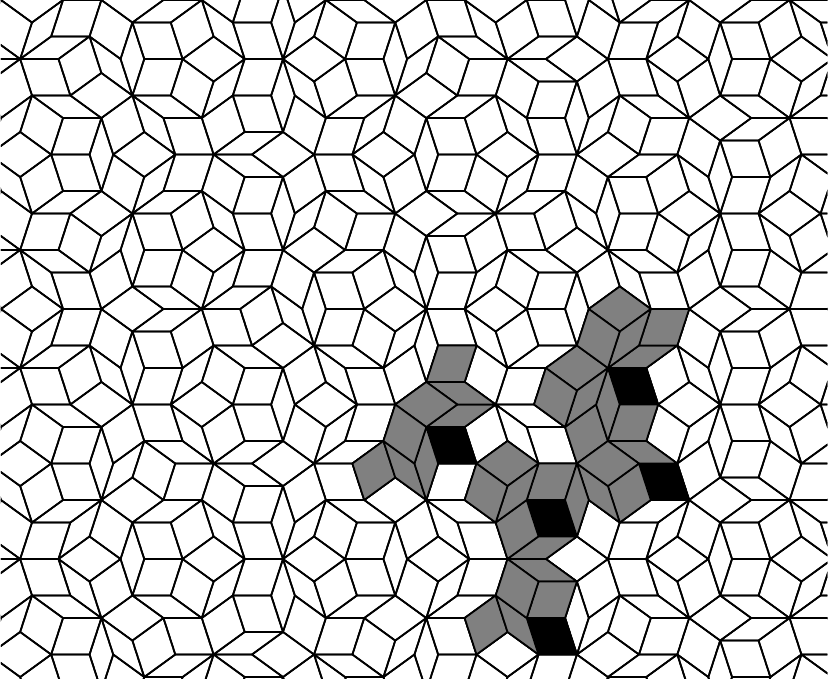}
    \includegraphics[width=0.23\linewidth]{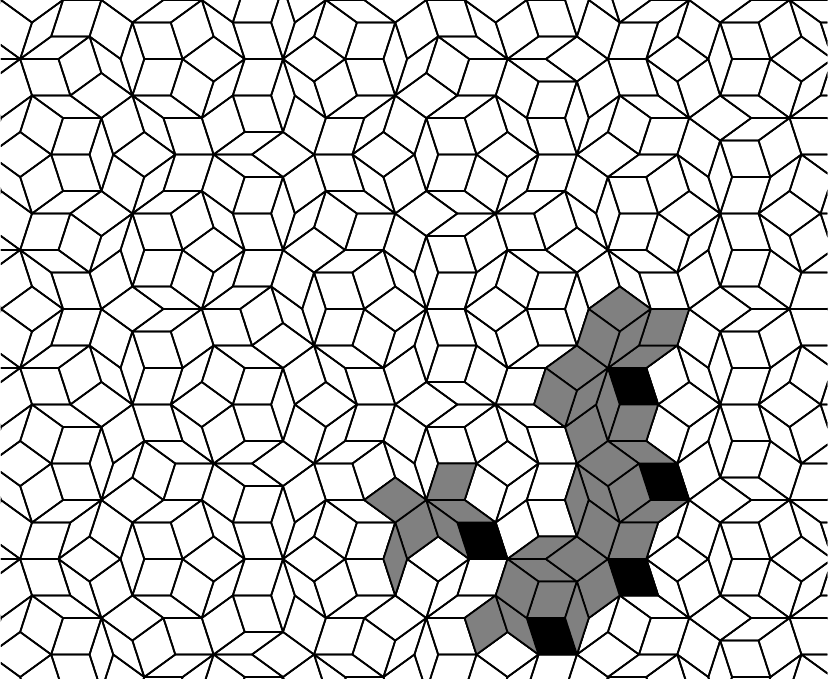}
    \includegraphics[width=0.23\linewidth]{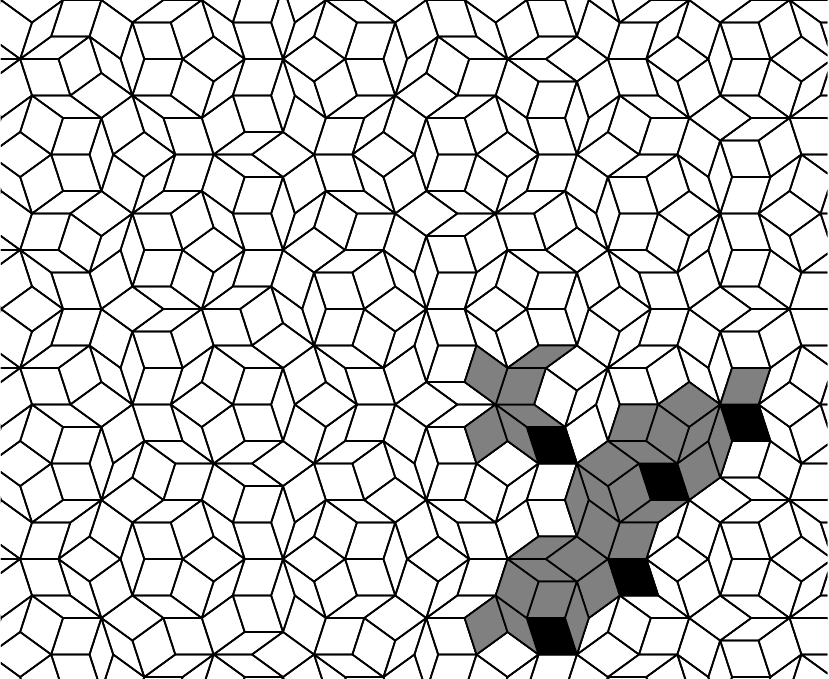} \vspace{.1cm} \\
  \end{center}
  \caption{Sixteen generations of a ``glider'' as it passes over a Penrose tiling, read from left-to-right, top-to-bottom.  Here we are using Game Variant (\ref{variant2}); the living dominant tiles are black, and the supporting tiles associated to living dominant tiles are gray.  The configurations in each column of the figure represent the glider in the same position (relative to the grid of dominant tiles).  The different configurations that represent the glider in the same position are due to the local structure of the tiling; a glider will assume identical configurations at times, but that timing is, itself, nonperiodic.}
  \label{f:PenroseGlider}
\end{figure}

\begin{enumerate}
\item \label{variant1}
We pick a pair of grid indices $(i,j)$, play the Game of Life on the lattice consisting of all $(i,j)$ type tiles, and declare all tiles that are not type $(i,j)$ to be ``dead."

\item \label{variant2} 
As in variant (i), we pick a pair of grid indices $(i,j)$ and play the Game of Life on the lattice consisting of all type $(i,j)$ tiles.  However, instead of declaring tiles of other types to be dead, we associate each tile that is not type $(i,j)$ to a nearby type $(i,j)$ tile.  Section \S \ref{ss:supporting} describes a canonical way to associate every tile that is not type $(i,j)$ to a nearby type $(i,j)$ tile.  We call the type $(i,j)$ tiles \emph{dominant}, and we call the set of non-dominant tiles associated to a dominant tile the \emph{supporting tiles}. 
We declare that a supporting tile is alive if and only if its associated dominant tile is alive.  
  (This is equivalent to specifying that the neighbors of a supporting tile are the eight dominant tiles that neighbor its associated dominant tile.)   (See Figure \ref{f:PenroseGlider}.)

\item In a regular tiling arising from a $n$-multigrid, there are ${n \choose 2}$ types of tiles, and each tile is of a unique type.  We simultaneously play a different copy of the Game of Life on each lattice consisting of all tiles of a given type. These copies do not interact.

\item As in variant (iii), we play a different copy of the Game of Life on each lattice consisting of all tiles of a given type, but on a singular tiling.  In this case, copies of the Game of Life played on different sets of tiles interact.  Tiles that belong to more than one lattice serve as channels of communication between the copies of the Game of Life played of these lattices.  (See Section \S \ref{ss:nonregular}.)  A different local rule may be required for these tiles, as they have more than eight neighbors.  
\end{enumerate} 

\section{The original Game of Life}

Let $X$ be a tiling of the plane, and let $T$ be the set of tiles in $X$.  
A cellular automaton played on $X$ with a set of states $\mathcal{A} =\{a_1,\dots,a_n\}$ is a map $f:T^{\mathcal{A}} \rightarrow T^{\mathcal{A}}$ such that 
\begin{enumerate}
\item each tile $t \in T$ has a ``neighborhood" $N(t) \subset T$ consisting of finitely many ``nearby" tiles
\item there is a ``local rule" map, $r$, defined on the set of all configurations of all neighborhoods and taking values in $\mathcal{A}$, so that for all configurations $c \in T^{\mathcal{A}}$ and all $t \in T$, $\pi_t(f(c))=r(\pi_{N(t)}(c))$, where $\pi_i$ is the restriction to the $i^{\textrm{th}}$ coordinate.  
\end{enumerate}
The space $T^{\mathcal{A}}$ is the phase space, or set of all possible configurations for the cellular automaton; a point $c \in T^{\mathcal{A}}$ is an assignment of one state in $\mathcal{A}$ to each tile $t \in T$.   Condition (ii) says that, regardless of the global configuration $c$, the next state of each tile $t$ is can be computed from the current states of the tiles in the neighborhood $N(t)$ using the local rule $r$.  

In Conway's original Game of Life (\cite{Gardner}), $T$ is the standard tiling of the plane by isometric squares, and $\mathcal{A} = \{0,1\}$.  Tiles in state $1$ are said to be  ``alive" or ``living," and tiles in state $0$ are said to be ``nonliving" or ``dead."  For each tile $t$, $N(t)$ is the eight tiles which neighbor $t$ (i.e. are vertically, horizontally, or diagonally adjacent to $t$).  The local rule $r$ is
 
 $$r(t) =   \begin{cases} 
1 \hspace{.5 cm} \textrm{ if } t \textrm{ is living and precisely 2 or 3 tiles in } N(t) \textrm{ are alive} \\ 
1 \hspace{.5cm} \textrm{ if } t \textrm{ is dead and precisely 3 tiles in } N(t) \textrm{ are alive} \\
0 \hspace{.5cm}  \textrm{ otherwise}
 \end{cases}$$
 
 Invented in 1970 by John Conway and popularized by Martin Gardner in an article in Scientific American (\cite{Gardner}), the Game of Life exhibits many interesting local configurations, including gliders, glider guns, spaceships, and replicators (cite).  It was shown to support universal computation, with the proof utilizing streams of gliders to transmit signals (\cite{WinningWays}). Dave Greene found the first replicator in the Game of Life in 2013 (\cite{Greene}).  
 
\section{The multigrid method} \label{s:multigrids}

In this section, we review the multigrid method (introduced by de Bruijn in \cite{deBruijn}) and use it to justify our assertion that the set of all ribbons passing through all type $(i,j)$ tiles forms a square grid graph.  

\subsection{General multigrids}
The multigrid method is a method of constructing quasiperiodic tilings.  A \emph{grid} is an infinite family of equally spaced, parallel lines in the plane.  (Note that his use of the word ``grid" differs from that of the ``square grid graph" of Section \S \ref{s:statementofresults} and Figure \ref{f:ribbons}.)  An \emph{$n$-multigrid}, $n \geq 2$, sometimes called just a \emph{multigrid}, is the union of $n$ grids in the plane, with no two grids parallel to the same vector.  A multigrid (or the associated tiling) is said to be \emph{regular} if no point is the intersection of more than 2 lines; otherwise it is \emph{singular}.  

A multigrid determines a tiling of the plane (see \S \ref{ss:Pentagrids}).  The tiling consists of one tile for each intersection of two (or more) lines in the multigrid. If precisely $2$ lines of the multigrid intersect at a point $v$, the tile corresponding to $v$ is a parallelogram whose edges are perpendicular to the lines intersecting at $v$. In the case of a singular multigrid, if $k\geq2$ lines of the multigrid intersect at $v$, the associated tile has one edge for each of the $2k$ half-lines incident to $v$, and the edges are perpendicular to the associated half-lines and arranged in the same order. 

Consequently, all tiles that correspond to intersections of a given line, say $L$, with other lines in a multigrid have opposite edges that are perpendicular to $L$ (and hence parallel).  The line $L$ corresponds in the tiling to a ribbon of tiles that all have a pair of edges that are perpendicular to $L$.   Two grids in the multigraph correspond in the tiling to a ``square grid graph" of ribbons, and the intersection points of these two grids correspond to the set of tiles that have edges perpendicular to the lines in both of these grids (see Figure \ref{f:ribbons}).

\subsection{Constructing Penrose tilings from pentagrids} \label{ss:Pentagrids}

For concreteness, we describe how to construct the tiling associated to a  $5$-multigrid (called  pentagrid) in which the grids are parallel to the vectors $1,\zeta^1,\zeta^2,\zeta^3,\zeta^4$, where $\zeta = e^{\frac{2\pi i}{5}}$.  This is the case described by de Bruijn and which gives rise to Penrose tilings.  An analogous process defines the tiling associated to a general $n$-multigrid (see, e.g. \cite{Senechal, deBruijn}). 

First, we define the $5$-multigrid, which in this case is also called a \emph{pentagrid}.  Fix real numbers $\gamma_0,\dots,\gamma_4$.  For $j= 0,...,4$, define the $j^{\textrm{th}}$ grid to be the set 
$$\{z \in \mathbb{C} \ |\textrm{ Re}(z\zeta^{-j}) + \gamma_j \in \mathbb{Z} \},$$ which consists of 
of equally spaced parallel lines.
The \emph{pentagrid}
determined by $\gamma_0,...,\gamma_4$ is the union of grids 0 to 4.  

We now define a map $K:\mathbb{C} \rightarrow \mathbb{Z}^5$ that describes the coordinate of points in the plane relative to each of the five grids.  Namely, $K$ associates to each point  $z \in \mathbb{C}$  
the point $K(z) \in \mathbb{Z}^5$ whose $j^{\mathrm{th}}$
coordinate is given by $$K_j(z) = \lceil \textrm{Re} (z\zeta^{-j})+\gamma_j
\rceil.$$
Define the map $\phi:\mathbb{Z}^5 \rightarrow
\mathbb{C}$ by 
$$\phi(\vec{k}) = \vec{k} \cdot
(1,\zeta^{1},\zeta^{2},\zeta^{3},\zeta^{4}).$$
The set of vertices of the associated tiling is the image of the plane $\mathbb{C}$ under $\phi \circ K$.  
As $z$ travels in a small circle around a point mentioned by $n \geq 2$ lines of the multigrid, the map $\phi \circ K(z)$ picks out a sequence of $2n$ points in $\mathbb{C}$; straight line segments connecting these $2n$ vertices (in cyclic order) are the edges of a tile in the tiling.  Each tile of the tiling comes from such an intersection of grid lines.  
De Bruijn proved \cite{deBruijn} that if the pentagrid determined by $\gamma_0,\dots,\gamma_4$ is regular, then the associated tiling is a Penrose tiling, and conversely, every Penrose tiling arises from such a pentagrid. 

A tile of type $(i,j)$, which comes from an intersection point of a line in the $i^\textrm{th}$ grid and a line in the $j^\textrm{th}$ grid, has edges parallel to $\zeta^i$ and $\zeta^j$.  If this intersection point is mentioned by precisely $n \geq 2$ lines of the grid, the tile will have $2n$ edges, of which two are parallel to $\zeta^i$ and two are parallel to $\zeta^j$.  If the tile is a quadrilateral, whether it is a thick rhomb or a thin rhomb depends on the parity of $i-j$.   A Penrose tiling exhibits ten different tile orientations (5~rotations of
both thin and thick tiles); these correspond to the $\binom{5}{2}$
ways that two of the five grids can intersect.  The intersection of a fixed line of grid $i$ with the other lines of the multigrid corresponds to a ribbon of tiles that all have edges parallel to $\zeta^i$. 

\begin{figure}[h]
  \begin{center}
    \includegraphics[width=.5\linewidth]{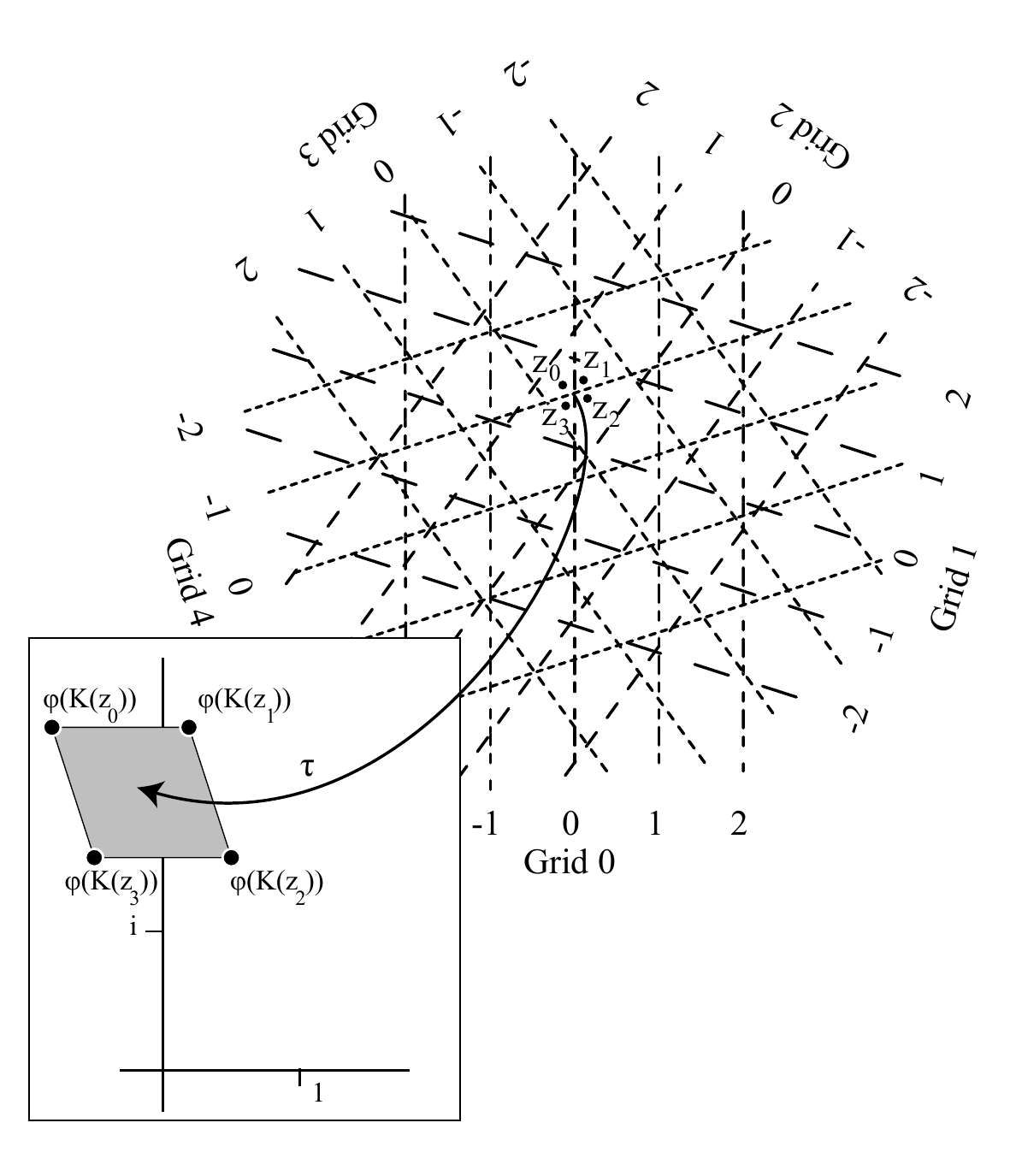}
  \end{center}
  \caption{The intersection, $z$, of line 0 of grid $i=0$ and line -1 of grid $j=4$.  The nearby points $z_0$, $z_1$, $z_2$, and $z_3$ are mapped by $K$ to four different points in $\mathbb{Z}^5$.  The images of these points under $\phi$ are the vertices of a tile in the associated tiling.}
  \label{f:Pentagrid}
  \end{figure}

\bigskip
\noindent \textbf{Example.}
Figure \ref{f:Pentagrid} shows a small section of a pentagrid.  Consider the
intersection of line $0$ of grid $0$ and line $-1$ of grid $4$ in the pentagrid; call
this point $z$.  In a small neighborhood of $z$, $K$ takes on four distinct
values:
$K(z_0)=(0,1,1,0,-1)$,
$K(z_1)=(1,1,1,0,-1)$,
$K(z_2)=(1,1,1,0,0)$,
and $K(z_3)=(0,1,1,0,0)$.
Thus, the point $z$ corresponds to the tile (a ``thick" rhomb) whose four vertices have coordinates\begin{align*}
\phi(K(z_0))&=\zeta^{1}+\zeta^{2}-\zeta^{4},\\
\phi(K(z_1))&=1+\zeta^{1}+\zeta^{2}-\zeta^{4},\\
\phi(K(z_2))&=1+\zeta^{1}+\zeta^{2},~\mathrm{and}\\
\phi(K(z_3))&=\zeta^{1}+\zeta^{2}.
\end{align*} Since $\zeta^j$ has unit magnitude, we can see that the edges of the
tile all have unit length.  Since the coordinates of the four points differ by a single multiple of $1$ and/or $\zeta^4$, the edges of the tile are parallel to the vectors $1$ or $\zeta^4$ in $\mathbb{C}$.

\section{Notes on Game Variants}

\subsection{Variant (ii): dominant and supporting tiles} \label{ss:supporting}

Section \S \ref{s:multigrids} describes a bijection between the set of tiles (of a tiling arising from a regular multigrid $G$) and the set of points in the plane that are mentioned by more than one grid of $G$.  For each pair of grid indices $(i,j)$, the union of grids $i$ and $j$ determines a decomposition of the plane into parallelograms. To be a true decomposition of the plane (i.e. each point belongs to precisely one parallelgram), view each parallelogram to be a product of left-closed, right-open (or right-open, left-closed) intervals.  If a point $v$ is in the intersection of two (or more) grids and belongs to parallelogram $P$, define the tile corresponding to $v$ to be in the support of the type-$(i,j)$ tile associated to the top left (or any other consistent choice) corner of the parallelogram $P$. 

\subsection{Variant (iv): singular tilings and interacting copies of the Game of Life} \label{ss:nonregular}

Suppose precisely 3 lines of a multigrid intersect at a point $z$; let $i,j,k$ be the indices of the three grids.  The tile associated to $z$ has six sides, is of types $(i,j)$, $(j,k)$, and $(i,k)$, and belongs to the three corresponding lattices of tiles.  Thus, this one tile may have as many as 24 neighbors -- eight neighbors in each of three lattices.  We may wish to impose a different local rule for tiles with more than eight neighbors.  The state of this tile impacts the evolution of each of the three ``copies" of the Game of Life.

In general, a tile that represents the intersection of $n \geq 3$ lines in a multigrid is of ${n \choose 2}$ types and thus belongs to ${n \choose 2}$ lattices, each of which hosts a copy of the Game of Life.  If each of its neighboring tiles arise from the intersection of precisely two multigrid lines (i.e. is a parallelogram), the tile will have $8  \cdot {n \choose 2}$ neighbors.   Understanding the qualitative effects of different possible local rules for tiles of more than one type is beyond the scope of this work.  

In the case of Penrose tilings, given any 3 distinct edge directions, there is at most one tile that has edges in all 3 directions (as well as, possibly, other directions).  Thus, communication between the games played on the three associated lattices of tiles can occur only at this one location.  Such a tile results from the intersection of lines in 3 distinct grids of the pentagrid.  To see that there can be at most one such tile, we may assume without loss of generality that two of the grids are grid $0$ and grid $1$, corresponding to the vectors $\zeta^0$ and $\zeta^1$.  Let $M$ be an affine deformation of the complex plane that maps these two vectors to $1$ and $i$ in $\mathbb{C}$.  Denote by $u$ the image under $M$ of $\zeta^j$, for any fixed $j \in \{2,3,4\}$.  Since both the real and complex coordinates of $u$ are irrational, the three vectors $1$, $i$, and $u$ are rationally independent.  Therefore, at most one point is in the intersection of grids $0$, $1$, and $j$; this proves the assertion that a (nonregular) Penrose tiling has most one tile of type $(0,1,j)$.  

\section{Discussion}
In some ways, our approach may be seen to have skirted the problem of how to ``compute" in nonperiodic substrates.  However, the fundamental observation of this work is that nonperiodic tilings may have periodic features that can be harnessed for computation.  The collection of all tiles of type $(i,j)$ naturally forms a lattice, but tiles that occupy adjacent positions \emph{in the lattice} may not be adjacent in the tiling as a whole (they are, however, a bounded distance apart). Game variant (\ref{variant2}) may be interpreted as a way to address this issue; one may consider the union of a dominant tile together with its supporting tiles as a single giant tile, eliminating the dead space between dominant tiles.  While interaction between copies of the game of life played on different lattices can occur at tiles with more than four sides, (nonregular) Penrose tilings have at most one tile that belongs to any three distinct lattices; generalizations of multigrids in which lines of the grid are not required to be spaced equally could give rise to more tiles that are loci for communcation between the lattices.  It remains to be seen to what extent our approach can inform implementation of computation in real-world nonperiodic substrates.  

\bibliographystyle{plain}
\nocite{*}
\bibliography{PenroseBibliography}

\end{document}